\documentclass{ieee}

\usepackage{amsmath}
\usepackage{amsfonts}
\usepackage{xspace}
\usepackage{hyperref}
\usepackage{times}

\usepackage{notation-hein}

\newtheorem{theorem}{Theorem}
\newtheorem{lemma}[theorem]{Lemma}
\newtheorem{definition}[theorem]{Definition}

\newenvironment{proof}[1][Proof]{%
  \begin{trivlist}{}{\setlength{\topsep}{0cm}\setlength{\partopsep}{0cm}}
  \item \textbf{#1.\@}\hspace*{1ex}\ignorespaces}%
  {\phantom{.}~\hfill$\Box$\end{trivlist}}

\allowdisplaybreaks[2]
\title{Multiparty Quantum Coin Flipping} 
\author{Andris Ambainis\thanks{supported in part by NSF Grant DMS-0111298}\\
  IAS and U.\ of Latvia%
  \and 
  \addtocounter{footnote}{1}
  Harry Buhrman\thanks{supported in part by the EU fifth framework 
    projects QAIP, IST-1999-11234, and RESQ, IST-2001-37559, and a 
    NWO grant}\\
  CWI and U.\ of Amsterdam%
  \and 
  Yevgeniy Dodis\thanks{supported in part by the National Science Foundation under CAREER Award No. CCR-0133806 and Trusted Computing Grant No. CCR-0311095}\\
  New York University%
  \and  
  \addtocounter{footnote}{-2}Hein R\"ohrig\footnotemark\\
  U.\ of Calgary%
}

\date{April 6, 2004}

\begin{document}

\maketitle
\thispagestyle{empty}

\begin{abstract}
\noindent
We investigate coin-flipping protocols for multiple parties in a
  quantum broadcast setting: 

\begin{itemize}
\item
We propose and motivate a definition for quantum broadcast. Our model
of quantum broadcast channel is new.

\item 
We discovered that quantum broadcast is essentially a combination of
pairwise quantum channels and a classical broadcast channel. This is a
somewhat surprising conclusion, but helps us in both our lower and
upper bounds.

\item 
We provide {\em tight} upper and lower bounds on the optimal bias
$\epsilon$ of a coin which can be flipped by $k$ parties of which exactly
$g$ parties are honest: for {\em any} $1\le g\le k$, $\epsilon =
\frac{1}{2} - \Theta\left(\frac{g}{k}\right)$. 
\end{itemize}
Thus, as long as a constant fraction of the players are honest, they
can prevent the coin from being fixed with at least a constant
probability. 
This result stands in sharp contrast with the classical setting, where
no non-trivial coin-flipping is possible when $g \le \frac k 2$.
\end{abstract}

\section{Introduction}

\subsection{The problem}
Consider $k$ parties out of which at least $g\ge 1$ are honest and at
most $(k-g)$ are dishonest; which players are dishonest is fixed in
advance but unknown to the honest players. The players can communicate
over broadcast channels.  Initially they do not share randomness, but
they can privately flip coins; the probabilities below are with
respect to the private random coins. A coin-flipping protocol
establishes among the honest players a bit $b$ such that
\begin{itemize}
\item if all players are honest, $\Pr [ b = 0 ] = \Pr [ b = 1 ] =
  \frac1 2$
\item if at least $g$ players are honest, then $\Pr [ b = 0 ], \Pr
  [ b = 1 ] \le \frac1 2 + \epsilon$
\end{itemize}
$\epsilon$ is called the {\em bias}; a small bias implies that
colluding dishonest players cannot strongly influence the outcome of
the protocol. 
Players may abort the protocol.

\subsection{Related work}
Classically, if a (weak) majority of the players is bad then no bias
$< \frac 1 2$ can be achieved and hence no meaningful protocols
exist~\cite{saks:cf}. For example, if we only have two players and one
of them is dishonest, then no protocols with bias $< \frac 1 2$ exist. For a
minority of bad players, quite non-trivial protocols exist. For
example, Feige~\cite{feige:selection} elegantly showed that
$(\frac{1}{2} + \delta)$-fraction of good players can achieve bias
$\frac{1}{2} - \Omega(\delta^{1.65})$, while achieving bias better
than $\frac{1}{2} - \delta$ is impossible.

Allowing quantum bits (qubits) to be sent instead of classical bits
changes the situation dramatically. Surprisingly, already in the
two-party case coin flipping with bias $<\frac 1 2$ is possible, as was
first shown by Aharonov~\etal~\cite{atvy:escrow}. The best known bias is $\frac 1 4$ and this is
optimal for a special class of three-round protocols
\cite{ambainis:coin}; for a bias of $\epsilon$ at least $\Omega ( \log
\log \frac1\epsilon )$ rounds of communication are necessary
\cite{ambainis:coin}.  Kitaev (unpublished,
see~\cite{kitaev:coinflipping}) showed that in the two-party case no
bias smaller than $\frac 1 {\sqrt{2}} - \frac 1 2$ is possible.

A weak version of the coin-flipping problem is one in which we know in
advance that outcome 0 benefits Alice and outcome 1 benefits Bob. In
this case, we only need to bound the probabilities of a dishonest
Alice convincing Bob that the outcome is 0 and a dishonest Bob
convincing Alice that the outcome is 1. In the classical setting, a
standard argument shows that even weak coin flipping with a bias $<
\frac 1 2$ is impossible when a majority of the players is dishonest.  In
the quantum setting, this scenario was first studied under the name
{\em quantum gambling} \cite{goldenberg&al:qgambling}.  Subsequently,
Spekkens and Rudolph \cite{spekkens&rudolph:cheatsensitive} gave a
quantum protocol for weak coin flipping with bias $\frac 1 {\sqrt{2}} - \frac 1 2$
(\ie no party can achieve the {\em desired outcome} with probability
greater than $\frac 1 {\sqrt{2}}$). This was recently improved to $0.192$ by Mochon~\cite{mochon04:weakCoinFlipping}.
Notice that this is a better bias than in
the best strong coin flipping protocol of \cite{ambainis:coin}.

We also remark that Kitaev's lower bound for strong coin flipping does
not apply to weak coin flipping. Indeed, Mochon's protocol has a better 
bound than Kitaev's lower bound. Thus, weak protocols with arbitrarily
small $\epsilon>0$ may be possible.  The only known lower bounds for
weak coin flipping are that the protocol of
\cite{spekkens&rudolph:cheatsensitive} is optimal for a restricted
class of protocols \cite{ambainis:weakFlipLower} and that a protocol
must use at least $\Omega(\log\log \frac 1 \epsilon)$ rounds of
communication to achieve bias $\epsilon$ (shown in
\cite{ambainis:coin} for strong coin flipping but the proof also
applies to weak coin flipping).

\subsection{Our contribution}

In this paper, we focus on quantum coin flipping for more than two
players.  However, for our multiparty quantum protocols we will
first need a new two-party quantum protocol for \emph{coin flipping
with penalty for cheating}. In this problem, players can be heavily
penalized for cheating, which will allow us to achieve lower cheating
probability as a function of the penalty. This primitive and the
quantum protocol for it are presented in
Section~\ref{sec:penaltyCoinFlip}; they may be of independent
interest.

One way to classically model communication between more than two
parties is by a primitive called \define{broadcast}. When a player sends
a bit to the other players he broadcasts it to all the players at
once~\cite{benor-linial:cf}.  However, when we deal with qubits such a
broadcast channel is not possible since it requires to clone or copy
the qubit to be broadcast and cloning a qubit is not
possible~\cite{nocloning}.  In Section~\ref{sec:broadcast} we develop
a proper quantum version of the broadcast primitive, which generalizes
the classical broadcast. Somewhat surprisingly, we show that our
quantum broadcast channel is essentially as powerful as a combination
of pairwise quantum channels and a classical broadcast channel. This
could also be of independent interest.

Using this broadcast primitive we obtain our main result:
\begin{theorem}\label{thm:upper&lower}
For $k$ parties out of which $g$ are honest, the optimal achievable
bias is $(\frac{1}{2} - \Theta(\frac{g}{k}))$.
\end{theorem}
We prove Theorem~\ref{thm:upper&lower} by giving an efficient protocol
with bias $(\frac{1}{2} - \Omega(\frac{g}{k}))$ in
Section~\ref{sec:multipartyProtocols} and showing a lower bound of
$(\frac{1}{2} - \bigO(\frac{g}{k}))$ in
Section~\ref{sec:lowerBounds}. Our protocol builds upon our
two-party coin-flipping with penalties which we develop in
Section~\ref{sec:penaltyCoinFlip}, and the classical protocol of
Feige~\cite{feige:selection} which allows to reduce the number of
participants in the protocol without significantly changing the
fraction of good players present. Our lower bound extends the
lower bound of Kitaev~\cite{kitaev:coinflipping}. 

To summarize, we show that quantum coin flipping is significantly more
powerful than classical coin flipping. Moreover, we give {\em tight}
tradeoffs between the number of cheaters tolerated and the bias of the
resulting coin achievable by quantum coin-flipping protocols. We also
remark that the fact that we obtain tight bounds in the quantum
setting is somewhat surprising. For comparison, such tight bounds are
unknown for the classical setting.

In the remainder of the paper, we assume some familiarity with quantum
computing. We recommend the book of Nielsen and Chuang~\cite{nielsen&chuang:qc} for
background information on this topic.

\subsection{Semidefinite programming}

Some of our proofs make use of duality in semidefinite programming.
For a review of semidefinite programming, see
\eg~\cite{laurent02:_handb_discr_optim}. Semidefinite programming is a
generalization of linear programming. In addition to linear
constraints, semidefinite programs (SDPs) may have constraints that
require that a square matrix of variables is positive semidefinite,
\ie that is is symmetric and all its eigenvalues are nonnegative.

We make use of the following basic properties of semidefinite
matrices. Let $A$, $B$, and $C$ denote square matrices acting on some
linear space $\mathcal V$ and $\mathcal W \subseteq \mathcal V$ a
subspace. If $A$ is positive semidefinite, we write $A \ge 0$.  We
define $A \ge B :\Leftrightarrow A-B \ge 0$. Then
\begin{gather*}
  A \ge B \Leftrightarrow \forall \ket\psi : \bra \psi A \ket \psi \ge
  \bra \psi B \ket \psi
  \displaybreak[2] \\
  A=B+C \text{ and } C \ge 0 \Rightarrow A \ge B
  \displaybreak[2] \\
  A \ge B \Rightarrow \tr_{\mathcal W} ( A ) \ge \tr_{\mathcal W} ( B )
\end{gather*}
Here $\tr_{\mathcal W} (A)$ denotes the partial trace.

In the Lagrange-multiplier approach, a constrained optimization
problem (called the \emph{primal}\/ problem)
\[
\max_{x \ge 0} f(x) \text{ subject to } g(x) \le a \quad \text{for fixed } a>0
\]
is reformulated as an unconstrained optimization problem
\[
\max_x \inf_{\lambda \ge 0} f(x) - \lambda \cdot
(g(x)-a)
\enspace ,
\] 
which is bounded from above by the constrained optimization problem
(the \emph{dual}\/ problem)
\[
\min_{\lambda \ge 0}
\lambda \cdot a \text{ subject to } (f - \lambda \cdot g)(x) \le 0 \text{ for all } x \ge 0 \enspace .
\]
In linear programming, $(f - \lambda \cdot g) (x) \le 0$
for all $x \ge 0$ if and only if $f - \lambda \cdot g \le 0$.
Therefore the preceding optimization problem can be simplified to
\[
\min_{\lambda \ge 0} \lambda \cdot a \text{ subject to } f - \lambda \cdot g
\le 0 \enspace .
\]
The same construction applies to SDPs using matrices as variables and
$A \cdot B := \tr ( A\adjoint B)$. A feasible solution of the dual
yields an upper bound on the optimal value of the primal problem.
Strong duality (\ie that the optimal values coincide) does not hold in
general; however, we will not need this below.

\section{Two-party coin flipping with penalty for cheating}
\label{sec:penaltyCoinFlip}

We consider the following model for coin flipping.  We have two
parties: Alice and Bob, among whom at least one is assumed to be honest.
If no party is caught cheating, the winner gets 1 coin, the loser gets
0 coins.  If honest Alice catches dishonest Bob, Bob loses $v$ coins
but Alice wins 0 coins. Similarly, if honest Bob catches dishonest
Alice, she loses $v$ coins but Bob wins 0 coins.

\begin{theorem}\label{thm:penaltyCoinFlip}
If Alice (Bob) is honest, 
the expected win by dishonest Bob (Alice) is
at most $\frac{1}{2}+\frac{1}{\sqrt{v}}$, for $v \ge 4$.
\end{theorem}

\begin{proof}
The protocol is as follows.
Let $\delta=\frac{2}{\sqrt{v}}$.
For $a \in \01$, define
$\ket{\psi_{a}}=\sqrt{\delta}\ket{a}\ket{a}+\sqrt{1-\delta}\ket{2}\ket{2} \in \C^3 \tensor \C^3$.

\begin{enumerate}
\item
Alice picks $a\in\{0, 1\}$ uniformly at random, generates the state
$\ket{\psi_a}$ and sends the second register to Bob.
\item
Bob stores this state in a quantum memory,
picks $b\in\{0, 1\}$ uniformly at random and sends
$b$ to Alice. 
\item
Alice then sends $a$ and the first register to Bob and Bob
verifies if the joint state of the two registers is $\ket{\psi_{a}}$ by
measuring it in a basis consisting of $\ket{\psi_{a}}$ and
everything orthogonal to it.
If the test is passed, the result of coin flip is $a\oplus b$, otherwise Bob catches Alice cheating.
\end{enumerate}
Theorem~\ref{thm:penaltyCoinFlip} follows immediately from the
following two lemmas.
\end{proof}

\begin{lemma}
Bob cannot win with probability more than 
$\frac{1}{2}+\frac{1}{\sqrt{v}}$, thus his expected win is at most 
$\frac{1}{2}+\frac{1}{\sqrt{v}}$.
\end{lemma}

\begin{proof}
Let $\rho_a$ be the density matrix of the second register of $\ket{\psi_a}$.
Then, for the trace distance between $\rho_0$ and $\rho_1$ we have 
$\|\rho_0-\rho_1\|_t=2\delta$. 

The trace distance is a measure for the distinguishability of quantum 
states analogous to the total-variation distance of probability 
distributions; see \eg~\cite{akn:mixed}. 
In particular, the probability of Bob winning is at most
$\frac{1}{2}+\frac{\|\rho_0-\rho_1\|_t}{4}=
\frac{1}{2}+\frac{\delta}{2}=\frac{1}{2}+\frac{1}{\sqrt{v}}$.
\end{proof}

\begin{lemma}
Dishonest Alice's expected win is at most $\frac{1}{2}+\frac{1}{\sqrt{v}}$.
\end{lemma}

\begin{proof}
  Alice is trying to achieve $a\oplus b=0$, which is equivalent to
  $a=b$.  We describe the optimal strategy of Alice as a semidefinite
  program.  

  The variables are semidefinite matrices over subspaces of
  $\mathcal X \tensor \mathcal A \tensor \mathcal
  B$, where $\mathcal X$ is Alice's private storage, $\mathcal A \isomorphic \C^3$ holds
  the first qutrit of the state to be sent in the protocol and
  $\mathcal B \isomorphic \C^3$ holds the second qutrit. For $a,b \in \01$, let
  $\rho_{ba} \in \mathcal A \tensor \mathcal B$ denote the state that
  Bob has in the last round, when he has sent $b$ and Alice has sent
  $a$ (and some qutrit). For $b \in \01$, let $\rho_b \in \mathcal X
  \tensor \mathcal A \tensor \mathcal B$ denote the
  state before Alice decides on $a$. Finally, let
  $\rho_\textup{initial} \in \mathcal X \tensor
  \mathcal A \tensor \mathcal B$ denote the state that Alice prepares
  initially and of which she sends the $\mathcal B$ part to Bob.
  
  Then we have the following constraints. The initial state is an
  arbitrary density matrix:
  \begin{equation}
    \label{eq:alice:1}
    \tr ( \rho_\textup{initial} ) = 1
  \end{equation}
  When Alice learns $b$, she cannot touch $\mathcal B$ anymore, but
  she can apply an arbitrary unitary $U_b$ on $\mathcal X
  \tensor \mathcal A$ to store her choice $a$ in $\mathcal
  X$ and to prepare the $\mathcal A$ register in the desired state:
  \begin{equation}
    \label{eq:alice:2}
    \tr_{\mathcal X \mathcal A} ( \rho_\textup{initial} ) = 
    \tr_{\mathcal X \mathcal A} ( \rho_b )
    \qquad \text{for all } b \in \01
  \end{equation}
  She will then measure $\mathcal X$ register in the computational
  basis to obtain $a$. Therefore we have
  \begin{equation}
    \label{eq:alice:3}
    \tr_{\mathcal X} ( \rho_b ) = \rho_{b0} + \rho_{b1} 
    \qquad \text{for all } b \in \01
    \enspace .
  \end{equation}
  Note that this implies $\tr(\rho_{b0}) + \tr(\rho_{b1})= 1$, so that
  in general the $\rho_{ba}$ are not density matrices.
  
  Now Bob checks $\rho_{ba}$. This gives rise to the following
  objective function for Alice's optimal cheating strategy:
  \begin{multline}
    \label{eq:alice:4}
    \max \sum_{\beta \in \01} \sum_{\alpha \in \01} \Pr[ b=\beta]
    \Pr[a = \alpha | b = \beta] \cdot \\ \bigl( \delta_{\alpha\beta} \Pr[
      \rho_{\beta\alpha} \text{ passes} ] - v \Pr[ \rho_{\beta\alpha}
      \text{ fails} ] \bigr)
  \end{multline}
  Here the Kronecker-Delta $\delta_{\alpha\beta} = 1$ if and only if $\alpha =
  \beta$ measures whether Alice managed to get $a$ and $b$ to match.
  Maximizing \eqref{eq:alice:4} is equivalent to maximizing
  \begin{multline}
    \label{eq:alice:5}
    \max \sum_{\beta \in \01} \sum_{\alpha \in \01} \Pr[ b=\beta]
    \Pr[a = \alpha | b = \beta] \cdot \\ \Pr[
    \rho_{\beta\alpha} \text{ passes} ] \left( \delta_{\alpha\beta}  
      + v \right)  
    \enspace .
  \end{multline}
  Bob plays honestly, therefore $\Pr[b=0] = \Pr[b=1] = \frac12$.
  Moreover, $\Pr[a = \alpha | b = \beta] = \tr( \rho_{\beta,\alpha})$
  and
  \begin{equation*}
    \Pr[ \rho_{\beta\alpha} \text{ passes} ] = \tr \left( \proj{ \psi_\alpha } 
      \frac{ \rho_{\beta,\alpha} }{ \tr ( \rho_{\beta,\alpha} ) } \right)
    \enspace .
  \end{equation*}
  Hence, $\Pr[a = \alpha | b = \beta] \Pr[ \rho_{\beta\alpha} \text{
    passes} ] = \bra{ \psi_\alpha } \rho_{\beta\alpha} \ket{
    \psi_\alpha }$. Substituting this into \eqref{eq:alice:5} and discarding
  the constant factor $\frac12$ gives the final objective function
  \begin{equation}
    \label{eq:alice:7}
    \max \sum_{\beta \in \01} \sum_{\alpha \in \01} 
    \bra{ \psi_\alpha } \rho_{\beta\alpha} \ket{
      \psi_\alpha }
    \left( \delta_{\alpha\beta}  
      + v \right)      
    \enspace .
  \end{equation}
  We now proceed to constructing the dual of the SDP formed by the
  objective function \eqref{eq:alice:7} together with the constraints
  \eqref{eq:alice:1}, \eqref{eq:alice:2}, and \eqref{eq:alice:3}. The Lagrange ansatz is
  \begin{multline}
    \label{eq:alice:8}
    \max_{\mathcal P} \inf_{\mathcal D} 
    \sum_{a,b \in \01} \tr \left( (\delta_{ab} + v ) 
      \proj{ \psi_a } \rho_{ba} \right)
    \\
    + \sum_{b \in \01} \tr \left( L_b \left( \tr_{\mathcal X} 
        ( \rho_b) - \rho_{b0} - \rho_{b1} \right) \right)
    \\ 
    - \sum_{b \in \01} \tr \left( M_b \tr_{ \mathcal X \mathcal A} 
      (\rho_b - \rho_\textup{initial} ) \right)
    \\ 
    - \tr( \lambda ( \rho_\textup{initial} - \id ) )
  \end{multline}
  where $\mathcal P$ are the primal variables as before, \ie
  \begin{multline*}
    \mathcal P = \{ ( \rho_\textup{initial}, \rho_0, \rho_1, \rho_{00}, 
    \rho_{01}, \rho_{10}, \rho_{11} ) \mathrel : \\ \rho_\textup{initial}, 
    \rho_0, \rho_1 \in S(\mathcal X \tensor \mathcal A 
    \tensor \mathcal B), \\ \rho_{00}, 
    \rho_{01}, \rho_{10}, \rho_{11} \in S(\mathcal A \tensor \mathcal B) \}
  \end{multline*}
  and the dual variables (Lagrange multipliers) are
  \begin{multline*}
    \mathcal
    D = \{ (L_0, L_1, M_0, M_1, \lambda) \mathrel: \\
    L_0, L_1 \in H(\mathcal A \tensor
    \mathcal B), M_0, M_1 \in H(\mathcal B), \lambda \in \R \}
    \enspace .
  \end{multline*}
  Here $H( \mathcal V )$ and $S(\mathcal V)$ denote the Hermitian and
  semidefinite matrices, respectively, operating on the linear space
  $\mathcal V$.  Collecting the primal variables in \eqref{eq:alice:8}, we
  get for $\rho_\textup{initial}$
  \begin{equation*}
    \tr \left( \left( M_0 + M_1 - \lambda \id_{\mathcal B} \right) 
      \rho_\textup{initial} \right)
    \enspace .
  \end{equation*}
  For $\rho_b$, $b \in \01$, we obtain
  \begin{equation*}
    \tr \left( 
      \left( 
        L_b - (\id_{\mathcal A} \tensor M_b)
      \right) 
      \tr_{\mathcal X} ( \rho_b )
    \right)
    \enspace .
  \end{equation*}
  For $\rho_{ba}$, $a,b \in \01$, we obtain
  \begin{equation*}
    \tr \left(
      \left(
        - L_b +
        ( \delta_{ab} + v )
        \proj{ \psi_a } 
      \right)
      \rho_{ba}
    \right)
    \enspace .
  \end{equation*}
  The terms in \eqref{eq:alice:8} not involving primal variables are
  just $\lambda$.  Hence, the following dual SDP will give an upper
  bound on the optimal value of our primal SDP:
  \begin{align}
    \label{eq:alice:16}
    \textup{minimize } & \lambda \textup{ subject to} 
    \\
    M_0 + M_1 & \le \lambda \id_{\mathcal B}
    \\
    L_b & \le \id_\mathcal{A} \tensor M_b  \textup{\ \ for all } b \in \01
    \\
    ( v + \delta_{ab}) \proj{ \psi_a } & \le L_b \textup{\ \ for all } a,b \in \01
    \\
    (L_0, L_1, M_0, M_1, \lambda) & \in \mathcal D
  \end{align}
  We now construct a feasible solution for the dual SDP. We restrict
  our attention to $M_0$ and $M_1$ of the form
  \begin{equation*}
    M_0 = 
    \begin{pmatrix}
      m_0 \\ & m_1 \\ & & m_2
    \end{pmatrix}   
    \text{ and }
    M_1 = 
    \begin{pmatrix}
      m_1 \\ & m_0 \\ & & m_2
    \end{pmatrix}      
  \end{equation*}
  for some $m_0$, $m_1$, $m_2 \in \R$ with
  \begin{equation}
    \label{eq:alice:36}
    m_0 \ge 0 \;, \qquad m_1 \ge 0 \;, \qquad m_2 = \frac12 (m_0+m_1)
    \enspace .
  \end{equation}
  Moreover, we also impose the restriction $L_b = \id_{\mathcal A}
  \tensor M_b$ for $b \in \01$.  Since then $\lambda \ge m_0 + m_1$,
  our goal reduces to minimizing $m_0$ and $m_1$ subject to
  \begin{align}
    \label{eq:alice:17}
    L_0 - (v+1) \proj{ \psi_0 } & \ge 0 \\
    \label{eq:alice:18}
    L_0 - v \proj{ \psi_1 } & \ge 0 \\
    \label{eq:alice:19}
    L_1 - v \proj{ \psi_0 } & \ge 0 \\
    \label{eq:alice:20}
    L_1 - (v+1) \proj{ \psi_1 } & \ge 0
    \enspace .
  \end{align}
  Constraints~\eqref{eq:alice:17} and \eqref{eq:alice:20} are satisfied if
  \begin{align}
    \label{eq:alice:34}
    m_0 & \ge (v+1) \delta \\
    m_2 & \ge (v+1) ( 1-\delta ) \\
    m_0 m_2 & \ge (v+1) (1-\delta) m_0 + (v+1) \delta m_2
    \enspace .
  \end{align}
  Similarly, Constraints~\eqref{eq:alice:18} and \eqref{eq:alice:19} require that
  \begin{align}
    m_1 & \ge v \delta \\
    m_2 & \ge v( 1 - \delta ) \\
    \label{eq:alice:35}
    m_1 m_2 & \ge v (1-\delta) m_1 + v \delta m_2 
    \enspace .
  \end{align}
  A solution to the system \eqref{eq:alice:36},\eqref{eq:alice:34}-\eqref{eq:alice:35}
  is
  \begin{equation*}
    \setlength{\arraycolsep}{0cm}
  \begin{array}{lll}
    m_0 & \; = \frac12 ( 1 + v ) 
    \Bigl( & 2 - \delta ( 1 + 2v )  \\
    && +
    {\sqrt{4 - 4\delta  + 
          {( \delta  + 2\delta v ) }^2}}
    \Bigr) 
    \\
    m_1 & \multicolumn{2}{l}{\; = \frac12 v \Bigl( 2 + \delta  + 2\delta v - 
        {\sqrt{4 - 4\delta  + 
            {( \delta  + 2\delta v ) }^2}}
      \Bigr)   \enspace . }
  \end{array}
  \end{equation*}
  From this and the definition of $\delta$, we get that there is
  feasible solution of the dual SDP with
  \begin{align*}
    \lambda & = m_0 + m_1 \\
    & = 2v + \frac{-1 + {\sqrt{v}} - 2 v + 
      {\sqrt{1 - 2 {\sqrt{v}} + 5 v + 4 v^2}}}{ \sqrt{v}}
    \\
    & \le 2v + 1 + \frac1{4 \sqrt v}
    \enspace .
  \end{align*}
  From the earlier transformations of the primal objective function,
  it follows that the optimal expected payoff of Alice is bounded from
  above by $\frac12 \lambda -v \le \frac12 + \frac1{8\sqrt v}$.
\end{proof}

\section{The multiparty model}
\label{sec:broadcast}

\subsection{Adversaries}

In this work, we assume computationally unbounded adversaries.
However, they have to obey quantum mechanics and cannot read the
private memory of the honest players (but they can communicate
secretly with each other). Moreover, we assume that they can only
access the message space in between rounds or when according to the
protocol it is their turn to send a message.

\subsection{The broadcast channel}

A classical \mention{broadcast} channel allows one party to send a classical bit
to all the other players. In the quantum setting this would mean that
a qubit would be sent to all the other players. However, when there
are more than two players in total we would have to {\em clone} or
{\em copy} the qubit in order to send it to the other players. Even if
the sender knows a classical preparation of the state he wants to
send, we cannot allow him to prepare copies because he may be a
cheater and send different states to different parties. It is well
known that it is impossible to clone a qubit~\cite{nocloning}, because
cloning is not a unitary operation. This means that we will have to
take a slightly different approach. Quantum broadcast channels have
been studied in an information-theoretic context before
\cite{bcfjs:noncommuting,wilmink:dis} but not in the presence of
faulty or malicious parties.

Our quantum broadcast channel works as follows. Suppose there are $k$
players in total and that one player wants to broadcast a qubit that
is in the state $\alpha \ket 0 + \beta \ket 1$. What will happen is
that the channel will create the $k$-qubit state $\alpha \ket {0^k} +
\beta \ket {1^k}$ and send one of the $k$ qubits to each of the other
players.  The state $\alpha \ket {0^k} + \beta \ket {1^k}$ can be
easily created from $\alpha \ket 0 + \beta \ket 1$ by taking $k-1$
fresh qubits in the state $\ket {0^{k-1}}$. This joint state can be
written as $\alpha \ket{0^k} + \beta\ket{10^{k-1}}$. Next we flip the
last $k-1$ bits conditional on the first bit being a $1$, thus
obtaining the desired state $\alpha \ket {0^k} + \beta \ket{1^k}$.
This last operation can be implemented with a series of controlled-not
operations. Note that this state is not producing $k$ copies of the
original state, which would be the $k$-fold product state
$(\alpha\ket{0} + \beta\ket{1} )\tensor \ldots \tensor(\alpha\ket{0} +
\beta\ket{1})$.

\begin{theorem}\label{thm:emulate}
  In the following sense, a quantum broadcast channel between $k$
  parties is comparable to models where the parties have a classical
  broadcast channel and/or pairwise quantum channels:
  \begin{itemize}
  \item If all parties are honest:
    \begin{enumerate}
    \item\label{QBCbyQC} One use of the quantum broadcast channel can
      be simulated with $2(k-1)$ uses of pairwise quantum channels.
    \item\label{BCbyQBC} One use of a classical broadcast channel can
      be simulated with one use of the quantum broadcast channel.
    \item\label{QCbyQBC} One use of a pairwise quantum channel can be
      simulated by $k+1$ uses of the quantum broadcast channel.
    \end{enumerate}
  \item If all but one of the parties are dishonest, using one of the
    simulations above in place of the original communication primitive
    does not confer extra cheating power.
  \end{itemize}
\end{theorem}
\begin{proof}
  We first give the simulations and argue that they work in case all
  players are honest.
  \begin{enumerate}
  \item[\ref{QBCbyQC}.] The sender takes $k-1$ fresh qubits in state
    $\ket{0^k}$. He applies $k-1$ times CNOT where the subsystem to be
    broadcast is the control of the CNOT and the fresh qubits are the
    destination. He then sends each of the $k-1$ qubits via the
    pairwise quantum channels to the $k-1$ other parties. Each
    recipient $j$ flips a (private) classical random bit $r_j$ and if
    $r_j = 1$ performs a $\sigma_z$ phase flip on the received qubit.
    Here $\sigma_z =
    \begin{pmatrix}
      1 & 0 \\ 0 & -1
    \end{pmatrix}$ is the Pauli matrix that multiplies the 
    relative phase between the $\ket 0$ and the $\ket 1$ state by $-1$.
    He then sends $r_j$ back to the sender. The sender computes the
    parity of the $r_j$ and if it is odd, he performs a $\sigma_z$
    phase flip on his part of the broadcast state, thus restoring the
    correct relative phase. (This randomization is a countermeasure;
    its utility is explained below.)
  \item[\ref{BCbyQBC}.] When the sender wants to broadcast bit $b \in
    \01$, he uses the quantum broadcast channel on qubit $\ket b$.
    The recipients immediately measure their qubit in the
    computational basis to obtain the classical bit.
  \item[\ref{QCbyQBC}.] The quantum broadcast channel can be used to
    create an \mention{EPR pair} $\frac 1 {\sqrt{2}} (\ket {00} + \ket{11}) $
    between two players $P_i$ and $P_j$ with the assistance of the
    other $(k-2)$ players. $i$ and $j$ are determined by the protocol.
    
    First one player broadcasts the state $\frac 1 {\sqrt{2}} (\ket 0 + \ket 1)
    $, resulting in the $k$ qubit state $\ket \phi = \frac 1 {\sqrt{2}} (\ket
    {0^k} + \ket{1^k})$. Now one after 
    the other, the $k-2$ remaining players perform a Hadamard
    transformation on their qubit, measure it in the computational
    basis, and broadcast the classical result.  Next, if $P_i$
    receives a $1$ he applies a phase flip $\sigma_z$
    to his part of $\ket \phi$ ($P_j$ does
    nothing). After this operation, $\ket \phi$ will
    be an EPR state between $P_i$ and $P_j$ unentangled with the other
    $k-2$ parties. Using a shared EPR pair, a protocol called {\em
      teleportation} \cite{teleporting} can be used to simulate a
    private quantum channel between $P_i$ and $P_j$. Teleportation
    requires the transmission of two bits of classical information.
  \end{enumerate}
  For the case of all but one party being dishonest:
  \begin{enumerate}
  \item[\ref{QBCbyQC}.] If the sender is honest, the recipients obtain
    exactly the same subsystems as for the quantum broadcast channel.

    If one of the recipients is honest, he may receive an arbitrary
    quantum subsystem up to the randomized relative phase. However,
    exactly the same can be achieved with a quantum broadcast channel
    with $k-1$ cheating parties, who each perform a Hadamard
    transformation on their subsystem followed by a measurement in the
    computational basis.
  \item[\ref{BCbyQBC}.] If the sender is honest, all recipients obtain
    the same computational-basis state. 
    
    If one of the recipients is honest, he obtains a classical bit
    that is possibly randomized in case the dishonest sender does not
    broadcast a basis state. Since the sender can flip a coin himself,
    this does not give more cheating power.
  \item[\ref{QCbyQBC}.] If the sender is honest, we can assume without
    loss of generality that all cheating action is done after the EPR
    pair has been established, because the (merged) cheaters can
    easily recreate the original broadcast state and also compensate
    any phase flipping of the honest sender. However, after the EPR
    pair has been established, the sender unilaterally performs his
    part of the teleportation circuit and measurements and sends the
    two bits of classical information. So the most general cheating
    action is to apply a quantum operation after the reception of the
    two classical bits. Furthermore, we can even assume that the
    cheating action is done \emph{after} the correction circuit of
    teleportation (this is similar to the teleportation of quantum
    gates~\cite{GC:tele}) and, hence, amounts to cheating on a
    pairwise quantum channel.

    If one of the recipients is honest, the best the cheaters can aim
    for is to give an arbitrary quantum state to the honest recipient.
    This they can also achieve over a pairwise quantum channel.
  \end{enumerate}
\end{proof}

\section{Multiparty quantum protocols}
\label{sec:multipartyProtocols}

We will first consider the case of only one good player (\ie $g=1$) among $k$ players
and later extend our results to general $g$. 

\paragraph{One honest player.} We need to construct a protocol
with bias $\frac 1 2 - \Omega(\frac 1 k)$. Before proceeding to our actual
protocol, let us consider a simple protocol which trivially extends
the previous work in the two-party setting, but does not give us the
desired result.  The protocols is as follows: player $1$ flips a
random coin with player $2$, player $3$ flips a random coin with
player $4$ and so forth. In each pair, the player with the higher id
wins if the coin is $1$ and the one with the lower id if the coin is
$0$. The winners repeat the procedure. With each round of the
tournament, half of the remaining players are eliminated (if there is
an odd number of players at any moment, the one with the highest id
advances to the next round). When there are only two players left, the
coin they flip becomes the output of the protocol.  (Above we assume
we have private point-to-point quantum channels and a classical
broadcast channel, which is justified by Theorem~\ref{thm:emulate}.)

The elimination step can be implemented using the weak
two-party coin-flipping protocol by Spekkens and Rudolph
\cite{spekkens&rudolph:cheatsensitive} and the last round by the the
strong two-party coin-flipping protocol by
Ambainis~\cite{ambainis:coin}. If there is only one good player, the
probability that he makes it to the last round is
\[
\left( 1-\frac 1 {\sqrt{2}} \right)^{\lceil -1+\log k \rceil}
\enspace ;
\]
in this case, the
probability that the bad players can determine the output coin is
$\frac 3 4$. In case the good player gets eliminated, the bad players can
completely determine the coin. Hence, the overall probability that the
bad players can determine the coin is 
\[
1 - \frac{1}{4}
\left( 1-\frac{1}{\sqrt{2}} \right )^{\lceil -1+\log k \rceil} \le 1 -
\frac{1}{4k^{1.78}}
\enspace , 
\]
which corresponds to bias 
\[
\frac{1}{2} -
\Omega \left( \frac 1 {k^{1.78}} \right)
\enspace .
\]
  
Using the protocol by Mochon~\cite{mochon04:weakCoinFlipping} improves
the exponent slightly to $\approx 1.7$ but not all the way to $1$.
To improve the bound above to the desired value $\frac{1}{2} -
\Omega(\frac 1 k)$, we will use our coin-flipping protocol with penalty from
Section~\ref{sec:penaltyCoinFlip}. The idea is that in normal quantum
coin-flipping protocols for two parties, there are three outcomes for
a given player: ``win,'' ``lose,'' and ``abort.'' Looking at the
elimination tournament above, if an honest player loses a given coin
flipping round, he does not complain and the bad player wins the game.
However, if the honest player detects cheating, he can and will abort
the entire process, which corresponds to the failure of the dishonest
players to fix the coin. Of course, if there are few elimination rounds
left, bad players might be willing to risk the abort if they gain
significant benefits in winning the round. However, if the round
number is low, abort becomes prohibitively expensive: a dishonest
player might not be willing to risk it given there are plenty more
opportunities for the honest player to ``lose normally.'' Thus,
instead of regular two-party coin-flipping protocols, which do not
differentiate between losing and aborting, we can employ our protocol
for coin flipping with penalty, where the penalties are very high at
the original rounds, and eventually get lower towards the end of the
protocol. Specific penalties are chosen in a way which optimizes the
final bias we get, and allows us to achieve the desired bias $\frac 1 2 -
\Omega(\frac 1 k)$.
\begin{theorem}\label{thm:upperCoin}
  There is a strong quantum coin-flipping protocol for $k$ parties with
  bias at most $\frac 1 2 - \frac c k$ for some constant $c$, even with $(k-1)$
  bad parties.
\end{theorem}
\begin{proof}
We assume that $k = 2^n$ for some $n>0$, as it changes $c$ by at most
a constant factor. Let $Q_v$ be the maximum expected win in a
two-party protocol with penalty $v$.  Consider the following protocol
with $n$ rounds numbered $1$ to $n$.

At the beginning of round $i$, we have $2^{n+1-i}$ parties remaining.  We
divide them into pairs.  Each pair performs the two-party
coin-flipping protocol with penalty $(2^{n-i}-1)$.
The party with the lower id plays Alice and wins if the outcome is $0$; 
the party with the higher id plays Bob and wins if the outcome is
$1$. 
The winners proceed to round $(i+1)$.

At the beginning of round $(n-2)$, there are just $8$ parties remaining.
They perform three rounds of regular coin flipping
with no penalty using the protocol of \cite{ambainis:coin,kerenidis&nayak:weakCoinFlip}, 
in which no cheater can determine the coin with probability more than $\frac 3 4$.
This results in maximum probability of $\frac{63}{64}$ for fixing the
outcome when there is at least one good player among the $8$.  The result of the last round is the result of our $2^n$-party protocol.

Assume that the honest player has won the first $(n-j)$ coin flips and
advanced to round $(j+1)$.  Assume that the all other players
in round $(j+1)$ are dishonest.  Let $P_j$ be the maximum
probability with which $(2^{j}-1)$ dishonest players can fix the outcome in this case.

\begin{lemma}
\begin{equation}\label{eq:1}
1- P_{j}\geq (1-P_{j-1}) (1- Q_{2^{j-1}-1})
\end{equation}
\end{lemma} 

\begin{proof}
Let $p_{w}$, $p_{l}$, $p_c$ be the probabilities of the honest player
winning, losing and catching the other party cheating in the
round $(j+1)$ of the protocol.  Notice that $p_w+p_l+p_c=1$.
Then, the probability $P_j$ of $2^j-1$ dishonest parties fixing the
coin is at most $p_l + p_w P_{j-1}$. (If the honest player loses, they
win immediately. If he wins, they can still bias the coin in $j-1$
remaining rounds to probability at most $P_{j-1}$. If he catches his
opponent cheating, he exits the protocol and the dishonest players
have no more chances to cheat him.) Using $p_w = 1- p_l - p_c$, we
have
\begin{multline}\label{eq:2}
P_j \le p_l + p_w P_{j-1} = P_{j-1} +(1-P_{j-1})p_l - P_{j-1} p_c \\ =
P_{j-1}+ (1-P_{j-1}) \left( p_l-\frac{P_{j-1}}{1-P_{j-1}} p_c \right)
\end{multline} 
Next, notice that $P_{j-1}\geq 1-\frac{1}{2^{j-1}}$.  This is
because $2^{j-1}-1$ bad players could just play honestly when they
face the good player and fix the coin flip if two bad players meet in
the last round.  Therefore,
$\frac{P_{j-1}}{1-P_{j-1}}\geq 2^{j-1}-1$ and \eqref{eq:2} becomes
\begin{equation*}
P_j \leq P _{j-1} + (1 - P_{j-1}) (p_l-(2^{j-1}-1) p_c)
\end{equation*}
The term $p_l-(2^{j-1}-1) p_c$ is at most $Q_{2^{j-1}-1}$ because we
can interpret it as the expected payoff of the cheater that plays with
the honest player. Hence,
\begin{equation*}
P_j \leq P _{j-1} + (1 - P_{j-1}) Q_{2^{j-1}-1}
\enspace ,
\end{equation*}
which is equivalent to the desired \eqref{eq:1}.
\end{proof}

By applying the lemma inductively, we obtain
\begin{align*}
1-P_n & \geq (1-P_8) \prod_{j=4}^{n} (1- Q_{2^{j-1}-1}) \\
& \geq \frac{1}{64} \prod_{j=4}^{n} (1- Q_{2^{j-1}-1}) 
\enspace .
\end{align*}
Using the bound in
Theorem~\ref{thm:penaltyCoinFlip} we get
\begin{align*}
1 - P_n & \ge \frac{1}{64} \prod_{j=3}^{n-1} (1 - Q_{2^j-1}) \\ & \geq
\frac{1}{64} \prod_{j=3}^{n-1} \left( \frac{1}{2} -
\frac{1}{\sqrt{2^j-1}} \right) 
\\ 
& \ge \frac{1}{8\cdot 2^n}
\prod_{j=3}^{\infty} \left(1 - \frac{2}{\sqrt{2^j-1}}\right)
\enspace . 
\end{align*}
The last product is a positive constant.  Therefore,
for some constant $c>0$ we have $1-P_n\geq \frac{c}{2^n} =
\frac{c}{k}$, which means that the bias is at most $\frac{1}{2}
-\Omega(\frac{1}{k})$.
\end{proof}
\paragraph{Extending to many honest players.}
We can extend Theorem~\ref{thm:upperCoin} to any number $g\ge 1$ of
good players by using the classical lightest-bin protocol of
Feige~\cite{feige:selection}. 
This protocol allows us to reduce the total number of players until a
single good player is left without significantly changing the fraction
of good players, after which we can run the quantum protocol of
Theorem~\ref{thm:upperCoin} to get the desired result. Specifically,
Lemma~8 from \cite{feige:selection} implies that starting from $g =
\delta k$ good players out of $k$ players, the players can
(classically) select a sub-committee of $\bigO(\frac 1 \delta)
= \bigO(\frac k g)$ players containing at least one good player
with probability at least $\frac 1 2$.  Now this sub-committee can use the
quantum protocol of Theorem~\ref{thm:upperCoin} to flip a coin with
bias $\frac 1 2 -\Omega(\frac g k)$, provided it indeed contains at least one
honest player. But since the latter happens with probability at least
$\frac 1 2$, the final bias is at most $\frac 1 2 - \frac 1 2 \cdot \Omega(\frac g k) = \frac 1 2
- \Omega(\frac g k)$, as desired.

\section{Lower bound}
\label{sec:lowerBounds}

\subsection{The two-party bound}

For completeness and to facilitate the presentation of our
generalization, we reproduce here Kitaev's unpublished proof
\cite{kitaev:coinflipping} that any two-party strong quantum
coin-flipping protocol must have bias at least $\frac 1 {\sqrt{2}}$. The
model here is that the two parties communicate over a quantum channel.
\begin{definition}
  Let $\mathcal H := \mathcal A \tensor \mathcal M \tensor \mathcal B$
  denote the Hilbert space of the coin-flipping protocol composed of
  Alice's private space, the message space, and Bob's private space%
  . A $2N$-round two-party coin-flipping protocol is a tuple
  \begin{align*}
  (&U_{A,1}, \ldots, U_{A,N}, U_{B,1}, \ldots, U_{B,N}, \\ &\Pi_{A,0},
  \Pi_{A,1}, \Pi_{B,0}, \Pi_{B,1})
  \end{align*}
  where
  \begin{itemize}
  \item $U_{A,j}$ is a unitary operator on $\mathcal A \tensor \mathcal
    M$ for $j = 1, \ldots, N$,
  \item $U_{B,j}$ is a unitary operator on $\mathcal M \tensor \mathcal
    B$ for $j=1,\ldots,N$,
  \item $\Pi_{A,0}$ and $\Pi_{A,1}$ are projections from $\mathcal A$
    onto orthogonal subspaces of $\mathcal A$ (representing Alice's
    final measurements for outcome $0$ and $1$, respectively),
  \item $\Pi_{B,0}$ and $\Pi_{B,1}$ are projections from $\mathcal B$
    onto orthogonal subspaces of $\mathcal B$ (representing Bob's
    final measurements for outcome $0$ and $1$, respectively),
  \end{itemize}
  so that for 
  \begin{align*}
    \ket{ \psi_N } := \; & ( 1_{\mathcal A} \tensor U_{B,N} ) (
  U_{A,N} \tensor 1_{\mathcal B} ) ( 1_{\mathcal A} \tensor U_{B,N-1}
  ) \\ & ( U_{A,N-1} \tensor 1_{\mathcal B} ) \cdots ( 1_{\mathcal A}
  \tensor U_{B,1} ) ( U_{A,1} \tensor 1_{\mathcal B} )\ket 0
  \end{align*}
  holds
  \begin{align}
    (\Pi_{A,0} \tensor 1 _ {\mathcal M} \tensor 1 _ {\mathcal B} )
    \ket{ \psi_ N} & = (1 _ {\mathcal A} \tensor 1 _ {\mathcal M}
    \tensor \Pi_{B,0} ) \ket{ \psi_ N} \\
    (\Pi_{A,1} \tensor 1 _ {\mathcal M} \tensor 1 _ {\mathcal B} )
    \ket{ \psi_ N} & = (1 _ {\mathcal A} \tensor 1 _ {\mathcal M}
    \tensor \Pi_{B,1} ) \ket{ \psi_ N} \label{eq:honest1} \\
    \lb { (\Pi_{A,0} \tensor 1 _ {\mathcal M} \tensor 1 _ {\mathcal B}
      ) \ket{ \psi_ N} } & = \lb { (\Pi_{A,1} \tensor 1 _ {\mathcal M}
      \tensor 1 _ {\mathcal B} ) \ket{ \psi_ N} }
  \end{align}
\end{definition}
The first two conditions ensure that when Alice and Bob are honest,
they both get the same value for the coin and the third condition
guarantees that when Alice and Bob are honest, their coin is not
biased. A player aborts if her or his final measurement does not
produce outcome $0$ or $1$; of course, it is no restriction to delay
this action to the end of the protocol.

\begin{lemma}\label{lem:kitaev}
  Fix an arbitrary two-party quantum coin-flip\-ping protocol.  Let
  $p_{1*}$ and $p_{*1}$ denote the probability that Alice or Bob,
  respectively, can force the outcome of the protocol to be $1$ if the
  other party follows the protocol. Denote by $p_1$ the probability
  for outcome $1$ when there are no cheaters. 
  Then $p_{1*} p_{*1} \ge p_1$. 
\end{lemma}
Hence, if $p_1 = \frac 1 2$, then $\max \{ p_{1*}, p_{*1} \} \ge
\frac 1 {\sqrt{2}}$. To prove Lemma~\ref{lem:kitaev}, we construct the view of
a run of the protocol from an honest Alice's point of view, with Bob
wanting to bias the protocol towards $1$. The problem of optimizing
Bob's strategy is a semidefinite program.

\begin{lemma}\label{lem:kitaevPrimal}
  The optimal strategy of Bob trying to force outcome $1$ is the
  solution to the following SDP over the semidefinite matrices
  $\rho_{A,0}, \ldots, \rho_{A,N}$ operating on $\mathcal A \tensor
  \mathcal M$:
  \begin{align}
    \textup{maximize } & \; \phantom{=} \; \tr \left( ( \Pi_{A,1}
      \tensor 1_{\mathcal M} ) \rho_{A,N}
    \right) \textup{ subject to}\\
    \tr_{\mathcal M} ( \rho_{A,0} ) & = \mix { 0}{ 0 }_{\mathcal A } \label{eq:primalBase} \\
    \tr_{\mathcal M} ( \rho_{A,j} ) & = \tr_{\mathcal M} ( U_{A,j}
    \rho_{A,j-1} U_{A,j} \adjoint ) \ \ \ (1 \le j \le N) \label{eq:primalStep}
  \end{align}
\end{lemma}
\begin{proof}
  Alice starts with her private memory in state $\ket 0_{\mathcal A}$
  and we permit Bob to determine the $\mathcal M$ part of the initial state.
  Therefore all Alice knows is that initially, the space accessible to
  her is in state $\rho_{A,0}$ with $\tr_{\mathcal M} ( \rho_{A,0} ) =
  \mix 0 0_{\mathcal A}$. Alice sends the first message, transforming
  the state to $\rho'_{A,0} := U_{A,1} \rho_{A,0} U_{A,1} \adjoint$.
  Now Bob can do any unitary operation on $\mathcal M \tensor \mathcal
  B$ leading to $\rho_{A,1}$, so the only constraint is $\tr_{\mathcal
    M} ( \rho_{A,1} ) = \tr_{\mathcal M} ( \rho'_{A,0} )$. In the next round,
  honest Alice applies $U_{A,2}$, then Bob can do some operation that
  preserves the partial trace, and so forth. The probability for Alice
  outputting $1$ is $\tr ( ( \Pi_{A,1} \tensor 1_{\mathcal M} )
  \rho_{A,N} )$ because the final state for Alice is $\rho_{A,N}$ and
  she performs an orthogonal measurement on $\mathcal A$ with
  projections $\Pi_{A,0}$, $\Pi_{A,1}$, and $1_{\mathcal A} -
  \Pi_{A,0} - \Pi_{A,1}$ (which represents ``abort'').
\end{proof}

\begin{lemma}
  The dual SDP to the primal SDP in Lemma~\ref{lem:kitaevPrimal} is 
  \begin{align}
    \textup{minimize } & \; \phantom{=} \; \bra 0 Z_{A,0} \ket 0 \textup{ subject to} \label{eq:dualobj} \\
    Z_{A,j} \tensor 1_{\mathcal M} & \ge U_{A,j+1}^* (Z_{A,j+1} \tensor 1_{\mathcal M} ) U_{A,j+1} \label{eq:zstep} \\ 
    & \qquad\qquad\quad (\textup{for all } j : 0 \le j \le N-1) \nonumber \\
    Z_{A,N} & = \Pi_{A,1} \label{eq:zfinal}
  \end{align}
  over the Hermitian matrices $Z_{A,0}, \ldots Z_{A,N}$ operating
  on $\mathcal A$.
\end{lemma}
\begin{proof}
  We form the dual of
  the SDP in Lemma~\ref{lem:kitaevPrimal} as follows: it is equivalent
  to maximizing over the $\rho_{A,j}$ the minimum of
  \begin{multline}
    \label{eq:lagrangeAnsatz}
    \tr ( ( \Pi_{A,1} \tensor 1_{\mathcal M} ) \rho_{A,N} ) - \tr
    (Z_{A,0} ( \tr_{\mathcal M} ( \rho_{A,0} ) - \mix 0 0_{\mathcal M} ))
    \\ - \sum_{j=1}^{N} \tr ( Z_{A,j} \tr_{\mathcal M} ( \rho_{A,j} -
    U_{A,j} \rho_{A,j-1} U_{A,j} \adjoint))
  \end{multline}
  subject to the operators $Z_{A,j}$ on $\mathcal M$ being Hermitian
  (for $0 \le j \le N$). In the sum above, the terms containing
  $\rho_{A,j}$ for $0 \le j < N$ are
    \begin{multline*}
    - 
    \tr(Z_{A,j} \tr_{\mathcal M} ( \rho_{A,j}  )) \\ + \tr( Z_{A,j+1}
    \tr_{\mathcal M} (
    U_{A,j+1} \rho_{A,j} U_{A,j+1} \adjoint )) 
    \enspace ,
    \end{multline*}
    which equals
    \begin{multline*}
    \tr \Bigl( \bigl( -(Z_{A,j} \tensor 1_{\mathcal M}) + \\ 
        U_{A,j+1}\adjoint (Z_{A,j+1} \tensor 1_{\mathcal M} ) U_{A,j+1}
         \bigr) \rho_{A,j} \Bigr)
      \enspace .
    \end{multline*}
    Since this term must be non-positive, we arrive at the inequality
    (\ref{eq:zstep}).
  
  For $j=N$, we obtain the dual equality constraint~(\ref{eq:zfinal})
  and the dual objective function becomes the only summand of
  (\ref{eq:lagrangeAnsatz}) that does not involve any $\rho_{A,j}$.
\end{proof}

\begin{proof}[Proof of Lemma~\ref{lem:kitaev}]
  Let $Z_{A,j}$ and $Z_{B,j}$ ($0 \le j \le N$) denote the optimal
  solutions for the dual SDPs for a cheating Bob and a cheating Alice,
  respectively. For each $j$, $0 \le j \le N$, let 
  \begin{multline*}
    \ket{ \psi_j } :=
    ( 1_{\mathcal A} \tensor U_{B,j} ) ( U_{A,j} \tensor 1_{\mathcal B}
    ) \cdots \\ ( 1_{\mathcal A} \tensor U_{B,1} ) ( U_{A,1} \tensor
    1_{\mathcal B} )\ket 0  
  \end{multline*}
  denote the state of the protocol in round
  $j$ when both parties are honest. Let $F_j := \bra{\psi_j} (Z_{A,j}
  \tensor 1_{\mathcal M} \tensor Z_{B,j} ) \ket{\psi_j} $. We claim
  \begin{align}
    p_{1*} p_{*1} & = F_0 \label{eq:f0} \\
    F_j & \ge F_{j+1} \label{eq:fstep} \qquad (0 \le j < N) \\
    F_N &=   p_1 . \label{eq:fn}
  \end{align}
  Combining (\ref{eq:f0})--(\ref{eq:fn}), we obtain the desired
  $p_{1*} p_{*1} \ge p_1$. We now proceed to prove these claims.
  
  Note that the primal SDP from Lemma~\ref{lem:kitaevPrimal} is
  strictly feasible: Bob playing honestly yields a feasible solution
  that is strictly positive. The strong-duality theorem of
  semidefinite programming states that in this case, the optimal value
  of the primal and the dual SDPs are the same, and therefore $p_{1*}
  = \bra 0_{\mathcal A} Z_{A,0} \ket 0_{\mathcal A}$ and $p_{*1} =
  \bra 0_{\mathcal B} Z_{B,0} \ket 0_{\mathcal B}$ and
  \begin{align*}
    p_{1*} p_{*1} & = \bra 0_{\mathcal A} Z_{A,0} \ket 0_{\mathcal A}
    \cdot \bra 0_{\mathcal M} 1_{\mathcal M} \ket 0_{\mathcal M}
    \cdot \bra 0_{\mathcal B} Z_{B,0} \ket 0_{\mathcal B} \\
    & = \bra 0 ( Z_{A,0} \tensor 1_{\mathcal M} \tensor Z_{B,0} ) \ket
    0 = F_0.
  \end{align*}
  The inequalities~(\ref{eq:fstep}) hold because of the
  constraints~(\ref{eq:zstep}).  Equality~(\ref{eq:fn}) holds because
  by constraint~(\ref{eq:zfinal}) we have
  \begin{multline*}
  \bra \phi (Z_{A,N} \tensor 1_{\mathcal M} \tensor Z_{B,N} ) \ket
  \phi = \\ \lb{ (\Pi_{A,1} \tensor 1_{\mathcal M} \tensor 1_{\mathcal
      B}) (1_{\mathcal A} \tensor 1_{\mathcal M} \tensor \Pi_{B,1})
    \ket \phi }^2
  \end{multline*}
  for any $\ket \phi$; $\ket{\psi_N}$ is the final state of the
  protocol when both players are honest, so by
  equation~(\ref{eq:honest1}),
  \begin{multline*}
  \lb{ (\Pi_{A,1} \tensor 1_{\mathcal M} \tensor 1_{\mathcal B})
    (1_{\mathcal A} \tensor 1_{\mathcal M} \tensor \Pi_{B,1} ) \ket{
      \psi_N } }^2 \\ = \lb{ (\Pi_{A,1} \tensor 1_{\mathcal M} \tensor
    1_{\mathcal B} ) \ket{ \psi_N } }^2 = p_1 .
  \end{multline*}
\end{proof}

\subsection{More than two parties}

We will now extend Kitaev's lower bound to $k$ parties. As with the
upper bounds, we first start with a single honest player ($g=1$), and
then extend the result further to any $g$.

\begin{theorem}\label{thm:lowerCoin}
  Any strong quantum coin-flipping protocol for $k$ parties has bias at
  least 
  \[
  \frac 1 2 - \frac{\ln 2}k - \bigO \left( \frac1{k^2} \right)
  \] 
  if it has to deal with up to $(k-1)$ bad parties.
\end{theorem}
We consider the model of private pairwise quantum channels between the
parties; by Theorem~\ref{thm:emulate} the results immediately carry
over to the quantum broadcast channel.
\begin{definition}
  Let $\mathcal H := \mathcal A_1 \tensor \cdots \tensor \mathcal A_k
  \tensor \mathcal M$ denote the Hilbert space composed of the private
  spaces of $k$ parties and the message space. An $N$-round $k$-party
  coin-flipping protocol is a tuple
  \[
  (i_1, \ldots, i_N, U_1, \ldots, U_N, \Pi_{1,0}, \Pi_{1,1}, \ldots,
  \Pi_{k,0}, \Pi_{k,1})
  \]
  where
  \begin{itemize}
  \item $i_j$ with $1 \le i_j \le k$, $1 \le j \le N$, indicates whose
    turn it is to access the message space in round $j$,
  \item $U_j$ is a unitary operator on $\mathcal A_{i_j} \tensor
    \mathcal M$ for $j=1,\ldots,N$,
  \item for $1 \le i \le k$, $\Pi_{i,0}$ and $\Pi_{i,1}$ are
    projections from $\mathcal A_i$ to orthogonal subspaces of
    $\mathcal A_i$ (representing the measurement that party $i$ performs to determine outcome $0$ or $1$, respectively),
  \end{itemize}
  so that for $\ket{ \psi_N } := \tilde U_{i_N} \cdots \tilde U_{i_1}
  \ket 0$ and each pair $1 \le i < i' \le k$ and any $b \in \{ 0, 1
  \}$ holds
  \begin{align}
    \tilde \Pi_{i,b} \ket{ \psi_N} & = \tilde \Pi_{i',b} \ket{ \psi_N}
    \\
    \lb{ \tilde \Pi_{i,b} \ket{ \psi_N} } & = 
    \lb{  \tilde \Pi_{i,1-b} \ket{ \psi_N} } .
  \end{align}
  Here $\tilde U_j$ denotes the extension of $U_j$ to all of $\mathcal
  H$ that acts as identity on the tensor factors $\mathcal A_{i'}$ for
  $i' \ne i_j$; $\tilde \Pi_{i,b} := ( 1_{\mathcal A_1} \tensor \cdots
  \tensor 1_{\mathcal A_{i-1}} \tensor \Pi_{i,b} \tensor 1_{\mathcal
    A_{i+1}} \tensor \cdots \tensor 1_{\mathcal A_k} )$ is the
  extension of $\Pi_{i,b}$ to $\mathcal H$.
\end{definition}
\begin{lemma}\label{lem:kitaevExt}
  Fix an arbitrary quantum coin flipping protocol.  For $b \in \{ 0, 1
  \}$, let $p_b$ be the probability of outcome $b$ in case all players
  are honest. Let $p_{i,b}$ denote the probability that party $i$ can
  be convinced by the other parties that the outcome of the protocol
  is $b \in \{ 0, 1 \}$. Then
  \[
  p_{1,b} \cdot \ldots \cdot p_{k,b} \ge p_b
  \]
\end{lemma}
\begin{proof}[Proof of Lemma~\ref{lem:kitaevExt}]
  The optimal strategy for $k-1$ bad players trying to force outcome
  $1$ is the solution to the SDP from Lemma~\ref{lem:kitaevPrimal}
  where all the cheating players are merged into a single cheating
  player.

  Let $( Z_{i,j} )_{0 \le j \le N}$ denote the optimal solution for
  the dual SDP for good player $i$, $1 \le i \le k$.  For each $j$, $0
  \le j \le N$, let $\ket{ \psi_j } := \tilde U_j \cdots \tilde U_1
  \ket 0$ denote the state of the protocol in round $j$ when all
  parties are honest. Let $F_j := \bra{\psi_j} (Z_{1,j} \tensor \cdots
  \tensor Z_{k,j} \tensor 1_{\mathcal M} ) \ket{\psi_j} $. By a
  similar argument as in the proof of Lemma~\ref{lem:kitaev}, we have
  \begin{align}
    p_{1,1} \cdot \ldots \cdot p_{k,1} & = F_0 \label{eq:f0m} \\
    F_j & \ge F_{j+1} \label{eq:fstepm} \qquad (0 \le j < N) \\
    F_N &= p_1 \label{eq:fnm}
  \end{align}
  Hence, $p_{1,1} \cdot \ldots \cdot p_{k,1} \ge p_1$. Repeating the
  argument with the cheaters aiming for outcome 0 completes the proof.
\end{proof}

Theorem~\ref{thm:lowerCoin} is an immediate consequence.
\begin{proof}[Proof of Theorem~\ref{thm:lowerCoin}]
  Using the notation of Lemma~\ref{lem:kitaevExt}, we have $p_0 =
  \frac 1 2$. Let $q= \max_i p_{i,0}$ denote the maximum probability of any
  player forcing output $0$. By Lemma~\ref{lem:kitaevExt}, $q^k \ge
  p_{1,0} \cdot \ldots \cdot p_{k,0} \ge \frac 1 2$, from which follows that
  \[
  q \ge \left( \frac 1 2 \right)^{1/k} \ge 1 - \frac{\ln 2}k -
  \bigO \left( \frac1{k^2} \right) \enspace .
  \]
  By
  Theorem~\ref{thm:emulate} this result applies both to private
  pairwise quantum channels and the quantum broadcast channel.
\end{proof}

\paragraph{Extending to many honest players.}
Extension to any number of honest players follows almost immediately 
from Theorem~\ref{thm:lowerCoin}. Indeed, 
take any protocol $\Pi$ for $k$ parties tolerating $(k-g)$ 
cheaters.
Arbitrarily partition our players into $k' = \frac k g$ groups 
and view each each as one ``combined player.''  We get an induced
protocol $\Pi'$ with $k'$ ``super-players'' which achieves at least
the same bias $\epsilon$ as $\Pi$, and can tolerate up to $(k'-1)$ bad
players. By Theorem~\ref{thm:lowerCoin}, $\epsilon \ge \frac 1 2 -
\bigO(\frac1{k'}) = \frac12 - \bigO(\frac g k)$.

\section*{Acknowledgements}

We thank L.~Fortnow and J.-H.~Hoepman for useful discussions.

\bibliographystyle{hieee-hein}
\bibliography{preamble,body}


\end{document}